# Superconductivity at 43 K in a single C-C bond linked terphenyl

Ren-Shu Wang[1,2], Yun Gao[2]*, Zhong-Bing Huang[3]*, Xiao-Jia Chen[1]*

**Affiliations:**

[1]Center for High Pressure Science and Technology Advanced Research, Shanghai 201203, China

[2]School of Materials Science and Engineering, [3]Faculty of Physics and Electronic Technology, Hubei University, Wuhan 430062, China

*Correspondence to: X.J.C. (xjchen@hpstar.ac.cn) or Y.G. (gaoyun@hubu.edu.cn) or Z.B.H. (huangzb@hubu.edu.cn)

**Abstract**: Organic compounds are promising candidates to exhibit high temperature or room temperature superconductivity. However, the critical temperatures of organic superconductors are bounded to 38 K. By doping potassium into *p*-terphenyl consisting of C and H elements with three phenyl rings connected by single C-C bond in *para* position, we find that this material can have a superconducting phase with the critical temperature of 43 K. The superconducting parameters such as the critical fields, coherent length, and penetration depth are obtained for this superconductor. These findings open an encouraging window for the search of high temperature superconductors in chain link organic molecules.

**One Sentence Summary:** The present finding of 43 K superconductivity in potassium-doped *p*-terphenyl is very encouraging for the future search of new high temperature superconductors in organic molecules connected by single C-C bond in a chain structure.

**Main Text:**

Exploring new materials with high temperature superconductivity is one attractive field in modern condensed matter physics. Organic based compounds were suggested as candidates of high temperature or room temperature superconductors (1,2). This idea assumed that the interaction of electrons with bosons having much higher excitation energy than the phonon energy can result in a substantially higher critical temperature $T_c$. The first experimental evidence of superconductivity in organic metals was found in 1980 (*3*). Since then, numerous organic superconductors have been reported including electron donor molecules such as tetrathiafulvalene, bis-ethylenedithrio-tetrathiafulvalene, and tetramethyltetraselenafulvalene derivatives, as well as electron acceptor molecules such as tetracyanoquinodimethane, fullerides and polycyclic aromatic hydrocarbons (4-11). These progresses illuminated the potential of organic materials for high temperature superconductivity, though the record high $T_c$ among them was still limited to 38 K (*6*).

The pioneering work of Little (*1*) in the proposal of synthesizing an organic superconductor particularly emphasized the need of a chain structure connecting molecules for high temperature superconductivity. In such a chainlike compound, one-dimensionality merely serves to optimize the attractive electron-electron interaction mediated by the polarization of the side-chain for superconductivity. There have been many efforts in synthesizing such a kind of actual chainlike superconductors (*3*). The excitonic mechanism of Cooper pairing for organic superconductivity still calls for further experimental examination. Nevertheless, his proposal has boosted extensive search for organic conductors and superconductors (12,13).

In this work we choose a chain link molecule only consisting of C and H elements with three phenyl rings connected by single C-C bond in *para* position – *p*-terphenyl. By doping potassium, we find superconductivity at 43 K in this chainlike system. This is the highest $T_c$ achieved in environmentally friendly organic materials. The result reveals that chain link organic molecules indeed can exhibit high temperature superconductivity.

High-purity *p*-terphenyl (99.5%) and potassium metal (99%) were purchased from Sigma-Aldrich and Sinopharm Chemical Reagent, respectively. The former was purified by vacuum drying method and the latter was cut into small pieces. They were then mixed with a mole ratio of 3:1. The mixtures were heated at temperature 443-533 K for 24-168 hours in quartz tubes sealed under high vacuum ($1 \times 10^{-4}$ Pa). After annealing, the samples were moved to a glove box with the oxygen and moisture levels less than 1 ppm. Using this procedure, the samples turned to back. They were placed into several nonmagnetic capsules and sealed by germanium varnish for magnetization and Raman scattering measurements. Magnetization measurements were performed with a SQUID magnetometer (Quantum Design MPMS3) in the temperature range of 1.8-300 K. Raman scattering experiments were carried on at room temperature in an in-house system with Charge Coupled Device and Spectrometer from Princeton Instruments in a wavelength of 660 nm and power less than 1 mW to avoid possible damage of samples.

Superconductivity in potassium-doped *p*-terphenyl was revealed by the magnetization measurements. Figure 1 shows the dc magnetic susceptibility $\chi$ in the applied magnetic field of 100 Oe with the field cooling (FC) and zero-field cooling (ZFC) runs in the temperature range of 1.8-300 K. One can readily see that $\chi$ shows a sudden drop below 43 K in the ZFC run, while it exhibits a platform in the FC run. This shape of the magnetic susceptibility curve is consistent with the well-defined Meissner effect, supporting the occurrence of superconductivity in this sample.

The inset of Fig. 1 presents the magnetization hysteresis for applied fields up to 5000 Oe measured at temperature of 10 K. The hysteresis loop along the two opposite magnetic field directions provides evidence for the type-II superconductor. In the superconducting state at 10 K, the magnetization *M* decreases with applying magnetic field and then increases till saturating to zero after passing a dip at around 700 Oe. This indicates that the lower critical field $H_{c1}$ for this superconductor at 10 K is less than 700 Oe. The closeness of the hysteresis at 5000 Oe yields the upper critical field $H_{c2}$ at such a value at 10 K for this superconductor.

The obtained superconductivity in potassium-doped *p*-terphenyl was further supported by the evolution of the magnetic susceptibility-temperature curve with the applied magnetic fields (Fig. 2). The curve gradually shifts towards the lower temperatures with increasing magnetic field. For the magnetic fields less than 1000 Oe, all curves below the transition temperatures shift in a parallel way. Upon applying higher magnetic field, the $\chi$-$T$ curve moves upwards, and the transition temperature is reduced significantly. Above 2000 Oe, superconductivity is largely suppressed. It seems likely that 4000 Oe is not enough to suppress superconductivity. Besides the 43 K superconducting transition, the magnetic susceptibility in the ZFC run also shows a sudden drop at 120 K from the almost constant on the high temperature side. This implies that there also exists a possible 120 K superconducting phase in this sample.

The Meissner effect of this material is clearly demonstrated from the magnetization measurements as a function of magnetic field up to 1000 Oe in the ZFC run at various temperatures in the superconducting state. The results are summarized in Fig. 3. The linearity of the *M-H* curves at these selected temperatures is the fingerprint for the Meissner phase of this superconductor, yielding different $H_{c1}$'s for corresponding temperatures based on the high field deviation from the linear behavior. Figure 4 shows the temperature dependence of the critical fields of this 43 K superconductor. $H_{c1}$ increases with decreasing temperature, and the zero-temperature extrapolated value of $H_{c1}(0)$ is 750±3 Oe. Based on the $\chi$-$T$ curves measured at various magnetic fields shown in Fig. 2, $H_{c2}$ can be determined by two steps. First, $T_c$ for a given field is determined from the intercept of linear extrapolations from below and above the transitions. Second, the magnetic field at the corresponding $T_c$ is assigned as $H_{c2}(T_c)$. The obtained $H_{c2}(T)$ as a function of temperature is shown in the right panel of Fig. 4. Using the Werthamer-Helfand-Hohenberg formula (*14*), the zero-temperature $H_{c2}(0)$=9315±10 Oe is obtained.

From $H_{c2}(0)$ and $H_{c1}(0)$, we evaluate the zero-temperature superconducting London penetration depth $\lambda_L$ and Ginzburg-Landau coherence length $\zeta_{GL}$ using the expressions (*15*) $H_{c2}(0)=\Phi_0/2\pi\zeta_{GL}^2$ and $H_{c1}(0)=(\Phi_0/4\pi\lambda_L^2)\ln(\lambda_L/\zeta_{GL})$ with the flux quantum $\Phi_0=2.0678\times10^{-15}$ Wb. Substituting the obtained $H_{c2}(0)$ and $H_{c1}(0)$, we obtain $\zeta_{GL}$=194±3 Å and $\lambda_L$=334±5 Å for this new superconductor. Thus the Ginzburg-Landau parameter $\kappa=\lambda_L/\zeta_{GL}$=1.74 is obtained. The obtained $\zeta_{GL}$ is larger but $\lambda_L$ is smaller by an order of magnitude compared with the $C_{60}$ superconductors (*13*).

Raman spectroscopy, a powerful phase-sensitive tool, was employed to identify the superconducting phase of the synthesized samples. The measured Raman spectra for the pristine and potassium-doped *p*-terphenyl are displayed in Fig. 5. Five regions of Raman active modes in pristine *p*-terphenyl correspond to the lattice, C-C-C bending, C-H bending, C-C stretching, and C-H stretching vibrations, respectively. Obviously, the spectra of potassium-doped *p*-terphenyl are quite different as compared to the pristine material. A comparison to previous studies indicates that the Raman spectra and features match well with those in potassium-doped *p*-terphenyl synthesized with liquid ammonia (*16*). When *p*-terphenyl is doped by potassium, all low-frequency lattice modes are significantly suppressed and the C-C-C bending modes get weaker with some shifts, except a sharp peak at 581 cm$^{-1}$. The C-H stretching modes in the high-frequency region become invisible.

The big difference of the spectra between the pristine and doped samples is observed in the C-H bending and C-C stretching regions. The strong band at 1474 cm$^{-1}$ and the triple bands at 950, 985, and 1006 cm$^{-1}$ are all from the formation of bipolarons, driving the structural change of the molecule from the benzenoid to quinoid (*17*). The Raman active 1474 cm$^{-1}$ mode is due to the C-H bending of external rings. The appearance of this mode can be considered as the fingerprint for the newly formed bipolarons. The triple bands are from the in-plane ring bending (*17*). These new bands result from the loss of translational symmetry due to the bipolaron-induced structural change. The two bipolaronic bands at 1172 and 1218 cm$^{-1}$ correspond to the 1222 cm$^{-1}$ band in the pristine material, arising from a mode concentrated in the inner region of the molecule with no contribution from the terminal rings (*18*). Significant changes are also found for the strong bands of the pristine material at 1275, 1593, and 1605 cm$^{-1}$. Upon doping potassium, the inter-ring C-C stretching band at 1275 cm$^{-1}$ is corrected with the 1351 cm$^{-1}$ band (*19*). The observed upshift of 76 cm$^{-1}$ mainly reflects the length decrease of the C-C bonds between rings. Thus, the increase of π-bond order of the inter-ring C-C bond is expected. By contrast, the intra-ring C-C stretching bands at 1593 and 1605 cm$^{-1}$ merge to the bipolaronic bands centered at 1589 cm$^{-1}$ (17,20), with slight downward shifts in wavenumber. The downshifts are the result of the doping-induced increase of inclined C-C bond lengths within the rings. These Raman spectra provide rather clear evidence for the formation of bipolarons in our sample.

The 43 K superconducting transition is the highest yet reported for molecular superconductors. Such a high $T_c$ seems beyond the generally believed 40 K upper limit for electron-phonon mediated superconductivity. Bipolarons have been identified as the quasiparticles in potassium-doped *p*-terphenyl from the Raman spectroscopy measurements. The strong electron-lattice interaction not only results in the formation of bipolarons but also provides the natural driving force for the observed 43 K superconductivity. In this aspect, bipolaronic superconductivity (21,22) may have already been realized in this material. Recently, we (*23*) also found superconductivity at 7.3 K in potassium-doped *p*-terphenyl. Compared with 43 K superconductivity reported here, the difference is the concentration of bipolarons. The present sample with 43 K superconducting transition appears to be highly doped because of the enhanced bands of bipolarons.

The unique feature of the *p*-terphenyl-like molecules is the presence of the chain connecting the phenyl rings. Whether the chain structure as suggested by Little (*1*) is crucial for such high temperature superconductivity in our sample could be a future attractive topic, the present finding of 43 K superconductivity is very encouraging for the search of new superconductors in organic compounds composed of chainlike molecules.

24. This work was supported by the Natural Science Foundation of China. We thank Hai-Qing Lin and Ho-Kwang Mao for strong support and valuable discussion. X.J.C., Y.G., and Z.B.H. designed the project. R.S.W. and X.J.C. synthesized the samples and performed the Raman scattering and magnetization measurements. All authors analyzed the data and discussed the results. X.J.C. wrote the paper with the inputs of all authors.

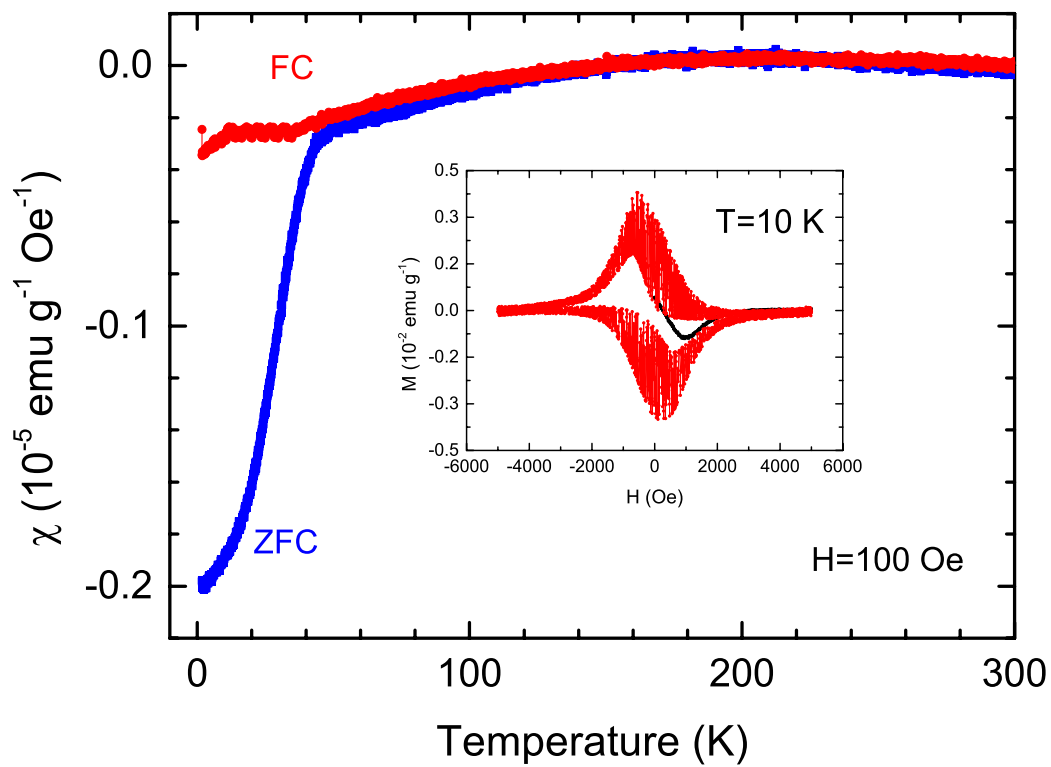

**Fig. 1**. The temperature dependence of the dc magnetic susceptibility χ for potassium-doped *p*-terphenyl in the applied magnetic field of 100 Oe with the field cooling (FC) and zero-field cooling (ZFC) runs in the temperature range of 1.8-300 K. Inset: The magnetization hysteresis with scanning magnetic field along two opposite directions up to 5000 Oe measured at temperature of 10 K.

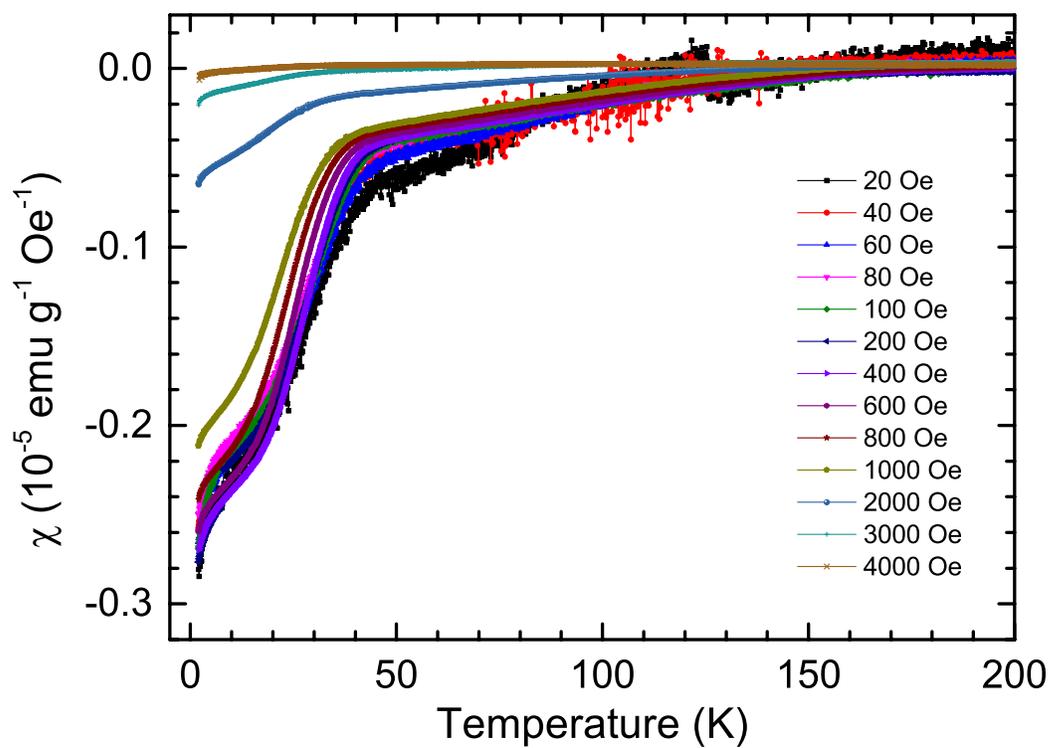

**Fig. 2.** The temperature dependence of the dc magnetic susceptibility of potassium-doped *p*-terphenyl measured at various magnetic fields up to 4000 Oe in the zero-field-cooling (ZFC) run.

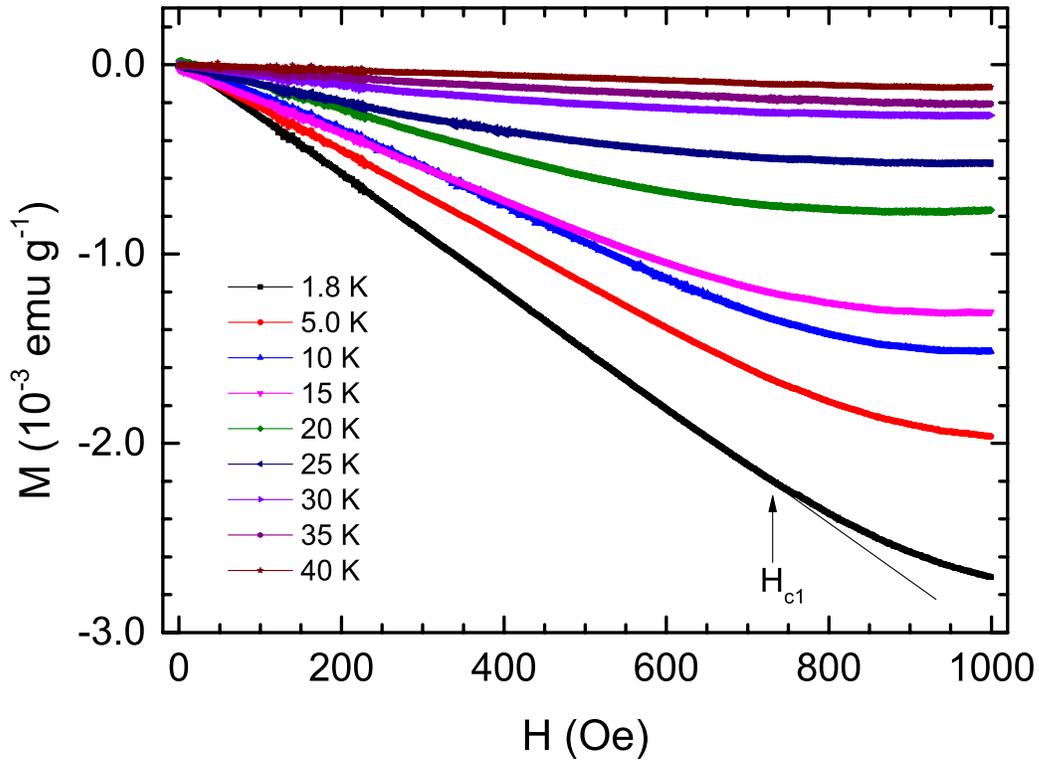

**Fig. 3.** The magnetic field dependence of the magnetization of potassium-doped *p*-terphenyl at various temperatures in the superconducting state measured in the zero-field-cooling (ZFC) run up to 1000 Oe. The lower critical field $H_{c1}$ is marked by an arrow defined by the deviation from the linear *M* vs *H* behavior.

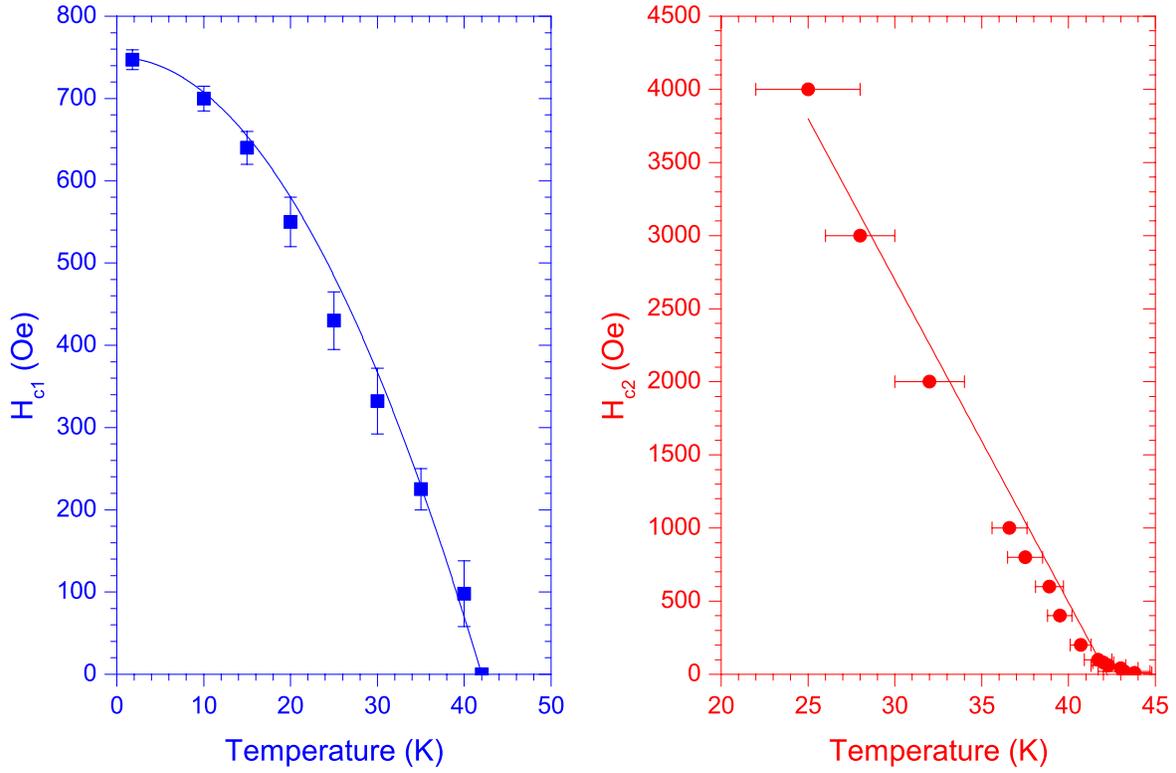

**Fig. 4.** Left, The temperature dependence of the lower critical field $H_{c1}(T)$. Error bars represent estimated uncertainty in determining $H_{c1}$. The solid line represents the empirical law $H_{c1}(T)/H_{c1}(0) = 1-(T/T_c)^2$. Right, The temperature dependence of the upper critical field $H_{c2}(T)$. The error bars represent the uncertainty in estimating the rounding of the transition. The line is the fitting to the Werthamer-Helfand-Hohenberg theory (*14*).

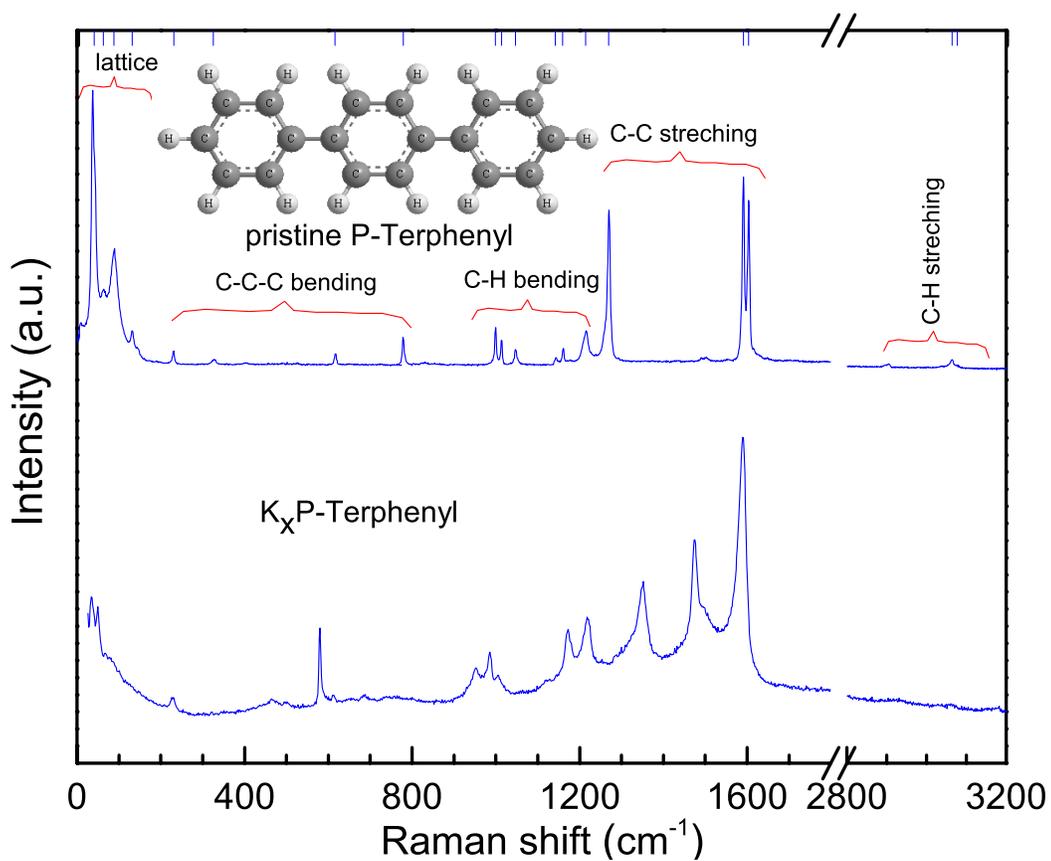

**Fig. 5.** Raman scattering spectra of pristine *p*-terphenyl and potassium-doped *p*-terphenyl collected at room temperature. Upper left presents the molecular structure of *p*-terphenyl. The sticks in the upper horizontal axis give the peak positions of the vibrational modes in pristine material.